\documentstyle[prd,aps]{revtex}
\begin{document}
 \input epsf
\draft
\renewcommand{\topfraction}{0.8}
%@twocolumnfalse\endcsname

\preprint{CITA-2000-05, hep-ph/0003018, March 2, 2000}
\title {\bf   On  the Theory of Fermionic Preheating  }
\author{Patrick B. Greene}
\address{Department of Physics, University of Toronto,
60 St. George Street, Toronto, ON M5S 1A7, Canada}
\author{Lev Kofman}
\address{ CITA, University of Toronto, 60 St. George Street,
Toronto, ON M5S 1A7, Canada}
\date { \today  }
\maketitle
\begin{abstract}

In  inflationary cosmology, the  particles constituting the Universe are
created after inflation due to their interaction
with moving inflaton field(s) in the process of preheating.
In the fermionic sector, the leading channel is out-of equilibrium
particle production in  the non-perturbative regime of parametric
 excitation, 
 which respects Pauli blocking
but  differs significantly from the perturbative expectation.
We develop theory of fermionic preheating coupling to the inflaton, 
 without and with 
 expansion of the universe, for light and  massive fermions, to
 calculate analytically the occupation number of  created fermions,
focusing on their spectra and time evolution.
In the case of large resonant parameter $q$ we extend 
for fermions the method of successive parabolic scattering, earlier  developed
  for bosonic preheating. 
In an  expanding universe parametric excitation of fermions is stochastic.
 Created  fermions  very quickly, within tens of inflaton oscillations, fill up
a sphere of radius $\simeq q^{1/4}$ in monetum space. 
We extend our formalism to the production  of superheavy fermions
and to `instant' fermion creation.
\end{abstract}
\pacs{PACS: 98.80.Cq  \hskip 2.5 cm CITA-2000-05 \hskip 2.5 cm
hep-ph/0003018}
% \vskip2pc]

\section{Introduction}

During cosmic inflation, it is assumed that entropy and
temperature associated with particles of matter are diluted to
practically zero values together with the number density of
particles. After inflation, the inflaton field $\phi$ oscillates
around the minimum of its effective potential $V(\phi)$. The energy of
inflaton oscillations is converted to the energy of newly
created particles of matter. Eventually particles of different
species are settled in a state of thermal equilibrium which marks
the beginning of the  conventional epoch of the hot Friedmann
radiation domination. The actual process of  particle creation
from the classical background inflaton oscillations occurs very
rapidly in the regime of parametric resonance~\cite{preh}  before the
thermal equilibrium will be settled. Particles are created
non-perturbatively  in the out-of-equilibrium state. The theory of
this process, {\it preheating}, is elaborated in details for the 
creation of bosons \cite{KLS97}. For bosons (denoted $\chi$), the leading
effect is the stimulated process of particle creation in the
regime of parametric resonance, where the number density of
created particles copiously increases with time as $n_{\chi} \sim
\exp{\int \mu dt}$. Soon, the backreaction of created $\chi$ particles
becomes important, so that the self-consistent dynamics of
interacting bose fields $\phi (t, {\vec x})$ and $\chi(t, {\vec
x})$, which can be treated classically,  can be revealed with
lattice simulations \cite{KT}.

At the beginning of the preheating investigation, a study  of fermion
production
 did not look very interesting relative to the bosonic
case. Indeed, the number density of fermions  $\psi$ is bounded by 
Pauli blocking. 
Therefore,  it 
  was not expected that fermions will
influence the dynamics of the inflaton and other scalar fields
(despite a numerical study \cite{boy95}  which down-played an observation 
that fermion production from an 
oscillating scalar is different from the perturbative prediction).

However,  it was  understood with surprise
\cite{BHP98,GK} that creation of fermions from the coherently
oscillating inflaton field occurs very differently than what the 
conventional perturbation theory of inflaton decay $\phi \to \bar
\psi \psi$ (say, due to a Yukawa coupling $h \bar \psi \phi\psi$)
  would predict, and this may have many interesting
cosmological applications.
It turns out that the occupation
number of fermions  very quickly, within about ten inflaton
oscillations, is saturated at the time-average value  of about
$n_{\psi} \sim 1/2$.
Moreover, in momentum space, fermions are
excited within a non-degenerate "Fermi sphere" of a
 large radius   $k \sim q^{1/4}m$, where $m$ is the frequency of
 inflaton oscillations, $q$ is the usual dimensionless
 parameter of parametric excitation, $q={{h^2 \phi^2_0} \over m^2}$.
Ironically, it may be parametric excitations of fermions that are 
responsible for the most important observable signatures or
observational constraints of preheating. Indeed, in some
inflationary models there is significant production of gravitinos
during the preheating stage \cite{gravitino}. Gravitinos are
cosmologically dangerous relics, for a mass of $\sim 100$ Gev
their abundance relative to that of  the relic photons cannot
exceed the bound $n_{3/2}/n_{\gamma} \leq 10^{-15}$. Theory of
gravitino production from preheating is rooted in the theory of
spin 1/2 fermionic preheating.  Another potentially important
application of fermionic preheating is a possibility to
produce superheavy fermions with a mass as large as $ \sim
10^{18}$ Gev from inflatons of mass $10^{13}$ Gev, as was
noticed in \cite{Cosmo,GPRT} and investigated in \cite{GPRT}.
Superheavy fermions may be interesting for the dark matter problem
and for the problem of ultra-high energy cosmic rays.
There are other interesting cosmological applications for 
the creation of fermions, e.g. the scenario of instant preheating \cite{inst} and
the creation of massive fermions during inflation \cite{infl}.
Recently, fermionic preheating in hybrid inflation for some range of parameters
 was thoroughly studied~\cite{hybrid}.
At present,
it is hard to say how important fermionic preheating will be in the
self-consistent non-linear  dynamics of bose- and fermi-fields
during preheating. Technical difficulty is to incorporate fermions
into lattice simulations alongside with bosons. However,  in
\cite{BHP98} it was reported  that even within the Hartree
approximation backreaction of fermions catalyzes bosonic
preheating.
 After all, for complete decay
of inflatons one needs a `three-legs' interaction, which is provided
naturally by the Yukawa coupling with fermions.

In this paper we develop the theory of fermionic preheating in an
expanding universe, following our short paper \cite{GK}, with an
emphasis on the analytic results. In particular, we will generalize for
 the fermionic
case some of the methods which we earlier developed for
bosonic preheating \cite{KLS97,GKLS97}, in particular, the method of parabolic
 scattering, which works for large values of $q$,
gives us an analytic
formula for the occupation number of fermions $n_k(t)$ as function of
time and momentum.
We extend this method for production of superheavy fermions from
a moving scalar field. 
 We begin Sect. 2
of  this paper with the Dirac equation and different VEVs for
fermions interacting with a time-dependent background scalar field.
 In Sect. 3 the creation of fermions without expansion of the universe
will be considered. 
We mostly will consider scalar
fields oscillating in a potential $V={{m^2 \phi^2}\over 2}$,
although the methods can easily be applied to others. 
We will give a semi-analytic treatment of the 
problem  based on some earlier results.
For the case $q \gg 1$ we
develop the method of successive parabolic scatterings.
 In Sect. 4 we take into account
expansion of the universe, when preheating of
 fermions acquires new qualitative
features. 
We extend the theory to describe  the production  of 
superheavy fermions. Our formalism also includes 
the case when fermions are created not from inflaton oscillations, but from
a single instance, when inflaton field crosses a certain level.

\section{\label{form} Formalism: Fermions Coupling with Background Scalar
in FRW metrics.}

	We will consider the creation of spin-$1 \over 2$ Dirac fermions $\psi$
by a homogeneous, oscillating scalar field $\phi$ in an expanding, flat FRW
universe.  Our strategy will be to solve the Heisenberg equation
of motion for the quantum $\psi$-field in the presence of the
classical backgrounds $\phi$ and $g_{\mu \nu}$.  Furthermore, we
will assume the energy-momentum of the $\phi$-field alone determines the
expansion rate of the Universe. The matter action we use
\begin{equation} \label{action}
S_{\rm M}
[\phi,\psi,e^{\alpha}_\mu] 
= \int{d^4 x \, e \left[ {1 \over 2} 
\partial_{\mu} \phi \partial^{\mu} \phi - V(\phi)
+ i \bar{\psi} \bar\gamma^\mu \stackrel{\rightarrow}{D}_\mu
\psi - (m_\psi + h\phi) \bar{\psi}\psi \right]} ~ ,
\end{equation} 
contains a simple Yukawa coupling between the scalar field with effective 
potential $V(\phi)$ and fermion with bare-mass $m_\psi$.  Here $\bar\gamma^\mu$
is a space-time dependent Dirac gamma matrix,
$e^{\alpha}_\mu$ is the vierbein with $e$ its determinant, and $D_\mu = 
\partial_\mu + {1 \over 4} \gamma_{\alpha \beta} \omega^{\alpha \beta}_\mu$
is the spin-${1 \over 2}$ covariant derrivative  with vierbein dependent
spin-connection, $\omega^{\alpha \beta}_\mu$.  The unbarred gammas are
standard, Minkowski space-time Dirac matrices and
$\gamma_{\alpha \beta} \equiv \gamma_{[\alpha}\gamma_{\beta]}$.
 
\subsection{\label{bkgdevol} Classical Background Fields}

We consider flat FRW metrics 
$ds^2=dt^2-a(t)^2d{\vec x}^2$ where
$a(t)$ is the
scale factor of the universe.  The Hubble parameter
$H \equiv {{\dot{a}} \over a}$ is  determined by the equation
$H^2 = {{8 \pi} \over {3 M^2_{\rm pl}}} 
\left({1 \over 2}(\dot{\phi})^2 + V(\phi)\right)$.
 The background homogeneous scalar field obeys the 
equation of motion
 $\ddot \phi + 3H \dot\phi + {{\partial V} \over {\partial \phi}} = 0$.
We will be most interested in fermion creation while
the $\phi$-field oscillates about a minimum of its potential. 
For illustration of the methods, we will consider
fermion creation in the context of chaotic inflation
scenarios with potentials of the form
$V(\phi) = {1 \over 2} m_\phi^2 \phi^2+ {\lambda \over 4}\phi^4 $.
After inflation, the $\phi$-field oscillates quasi-periodically with a
slowly decreasing amplitude due to the Hubble expansion.  The specific value of
the parameter $m_\phi$ or $\lambda$ (or some
combination thereof) will be dictated by cosmology.
We mostly will consider quadratic potential with $\lambda=0$.
In this paper we ignore backreaction of created particles. 

\subsection{\label{psievol} Quantum  Dirac Field}

	Variation of the action~(\ref{action})
with respect to $\bar{\psi}$ leads to the general relativistic
generalization of the Dirac equation
\begin{equation} \label{dirac_gen}
\left[i \bar\gamma^\mu D_\mu - (m_\psi + h\phi(t))\right]\psi(x) = 0 ~ .
\end{equation}
It can be shown that for an FRW space-time we have $\bar\gamma^0 = \gamma^0$
and $\bar\gamma^i = {1 \over {a(t)}}\gamma^i$, where the $\gamma^\alpha$'s
are standard, Minkowski space-time Dirac matrices.  Furthermore, spin
connection leads to 
\begin{equation}
{1 \over 4} \bar\gamma^\mu \gamma_{\alpha \beta} \omega^{\alpha \beta}_\mu
= {3 \over 2} \left({{\dot a} \over a}\right) \gamma^0  \ .
\end{equation}
Thus, in our case, the Dirac equation becomes
\begin{equation} \label{dirac_exp}
\left[ i\gamma^0 \partial_0 + i{1 \over a} {\vec\gamma \cdot \vec\nabla}
+ i {3 \over 2}\left({{\dot a} \over a}\right) \gamma^0 - (m_\psi + h\phi) 
\right] \psi(x) =0 \ .
\end{equation}
Only standard gamma matrices now appear.  Similarly,
$\bar\psi = \psi^\dagger \gamma^0$ is found to obey the conjugate of
Equation~(\ref{dirac_exp}).

In the Heisenberg representation of QFT, the Dirac
equation~(\ref{dirac_gen})~or~(\ref{dirac_exp}) becomes the equation of
motion for the $\psi(x)$ field operator.  $\psi(x)$ may be decomposed
into eigen-spinors
\begin{equation} \label{op_exp}
\psi({\bf  x},t)  = \sum_{s = \pm}
\int {d^3k \over{(2\pi)^{3}}} \ {\Bigl( \hat{a}_{k,s} {\bf u}_{k,s}(t)\,
e^{ +i{{\bf k} \cdot {\bf x}}} +  \hat{b}_{k,s}^{\dagger}
{\bf v}_{k,s}(t)\,  e^{ -i{{\bf k} \cdot {\bf x}}}  \Bigr)}  ,
\end{equation}
where ${\bf u}_{k,\pm}(t)$ is a positive-frequency eigenspinor of the Dirac
equation~(\ref{dirac_exp}) with helicity $\pm {1 \over 2}$ and
${\bf v}_{{k},\pm}(t) = {\cal C} {\bar {\bf u}}^T_{{k},\pm}(t)$ is its charge
conjugate.  Here ${\cal C}= i\gamma_0\gamma_2$ is the standard charge
conjugation matrix.  To construct these eigen-spinors we use
the ansatz
\begin{equation}
{\bf u}_{k,\pm}(t)\, e^{ +i{{\bf k} \cdot {\bf x}}}= a
\left[ -i\gamma^0 \partial_0 - i{1 \over a} {\vec\gamma \cdot \vec\nabla}
- i{3 \over 2}\left({{\dot a} \over a}\right) \gamma^0 - (m_\psi + h\phi) 
\right]
\chi_k(t) R_\pm ({\bf k}) e^{ +i{{\bf k} \cdot {\bf x}}} \ ,
\end{equation}
where $R_\pm ({\bf k})$ are eigenvectors of the helicity operator
${\bf k \cdot \Sigma}$ such that
${\bf k \cdot \Sigma} R_\pm ({\bf k}) = \pm 1$ and
$\gamma^0 R_\pm ({\bf k}) = +1$.  The time dependence of the
eigen-spinor is contained intirely in the
mode function $\chi_k(t)$.  Substituting this ansatz into
equation~(\ref{dirac_exp}) leads to the mode equation
\begin{equation}
\ddot{\chi}_k + 4{{\dot a} \over a}\dot{\chi}_k + \left[ {k^2 \over a^2}
+ m^2_{\rm eff} -i{{\dot {(a \, M_{\rm eff})}} \over a}
+{9 \over 4}\left({{\dot a} \over a}\right)^2
+ {3 \over 2}{ {\ddot a} \over a} \right] \chi_k = 0 \ ,
\end{equation}
where  $k^2 = \vert {\bf k} \vert^2$ and 
the   effective mass of the fermions is 
\begin{equation} \label{eff}
M_{\rm eff} \equiv (m_\psi + h\phi) \ .
\end{equation}
  The damping term in this
equation may be removed by defining a new mode function
$X_k(t) = a^2 \chi_k(t)$.  The mode equation becomes
\begin{equation} \label{mode_exp}
\ddot{X}_k + \left[ {k^2 \over a^2}
+ M^2_{\rm eff} -i{{ {(a \, M_{\rm eff})}} \over a}^.
+ \Delta(a) \right] X_k = 0 \ ,
\end{equation}
where
$\Delta(a) \equiv [{1 \over 4}\left({{\dot a} \over a}\right)^2
- {1 \over 2} \left({{\ddot a} \over a}\right)]$.  When $a \propto t^n$
such as in a matter or radiation dominated universe, $\Delta(a) \sim t^{-2}$
and can be neglected soon after inflation.

	Using the mode equation and ansatz, we find
\begin{equation} \label{uspin}
{\bf u}_{k,\pm}(t) = a^{-1} \left[ -i{\dot X}_k - i{1 \over 2}\left({{\dot a}
 \over a} \right) X_k
+ ({\bf \gamma \cdot k} - M_{\rm eff})X_k \right]R_\pm ({\bf k}) \ .
\end{equation}
Taking the charge conjugate of ${\bf u}_{k,\pm}$, we find
\begin{equation} \label{vspin}
{\bf v}_{k,\pm}(t) = a^{-1} \left[ i{\dot X}^*_k + i{1 \over 2}\left({{\dot a}
 \over a} \right) X^*_k
- ({\bf \gamma \cdot k} + M_{\rm eff})X^*_k \right] {\bar R}_\pm ({\bf k}) \ ,
\end{equation}
where ${\bar R}_\pm ({\bf k}) \equiv -i \gamma^2 R^*_\pm ({\bf k})$ is an
eigenvector of helicity such that
${\bf k \cdot \Sigma} {\bar R}_\pm ({\bf k}) = \pm 1$ and
$\gamma^0 {\bar R}_\pm ({\bf k}) = -1$.  

The energy-momentum tensor is obtained
from the ($\psi$ and $\bar\psi$ symmetrized) matter action by variation
with respect to the vierbein
\begin{equation} \label{emtensor}
T^{\mu \nu} = {i \over 2}
\left[\bar\psi\bar\gamma_{(\mu}\stackrel{\rightarrow}{D}_{\nu)}\psi
-\bar\psi \stackrel{\leftarrow}{D}_{(\mu}\bar\gamma_{\nu)}\psi \right] ~ ,
\end{equation}
and the Hamiltonian operator is
\begin{equation} \label{hamiltonian}
{\mathcal{H}}_D = \int{d^3x 
\left[i \psi^\dagger {\dot \psi} \right]} \ .
\end{equation}
In general, if this Hamiltonian is diagonal in the annihilation (and creation) 
operators $\hat{a}_{k,s}$ ($\hat{a}^\dagger_{k,s}$) and
$\hat{b}_{k,s}$ ($\hat{b}^\dagger_{k,s}$) at $t=0$, it will not be for
later times.  This is the signature of particle creation due to the time
dependent background. In order to determine the number of particles
produced, we perform a Bogliubov transformation on the creation and
annihilation operators so as to diagonalize the Hamiltonian at time $t$.

For the problem of particle creation quite often it is useful to represent the wave
function in 
the adiabatic ( semi-classical, WKB) form
\begin{equation}
X_k(t) ={\alpha_k} N_+ \,
 e^{-i\int_0^td t\, \Omega_k(t) } +
{\beta_k} N_- \, e^{+i\int_0^tdt\, \Omega_k(t) } \ ,
\label{adiab}
\end{equation}
where $N_{\pm}= (2\Omega_k(\Omega_k \pm M_{\rm eff}))^{-1/2}$ and 
the coefficients $\alpha_k$ and $\beta_k$  correspond to the 
coefficients of the Bogliubov transformation. 

Once the Bogliubov transformation is done,
we may write the comoving number density of particles
$ n_k(t)= \vert \beta_k \vert^2$
 in
a given spin state through the solutions of the mode equation~(\ref{mode_exp})
\begin{equation} \label{nk}
n_k(t) = a\left({{\Omega_k - M_{\rm eff}} \over {2\Omega_k}}\right)
\left[|\dot X_k|^2 + \Omega_k^2 |X_k|^2 - 2\Omega_k Im(X_k \dot X_k^*)\right]
\ ,
\end{equation}
where $M_{\rm eff} \equiv (m_\psi + h\phi)$ as before and $\Omega_k^2
\equiv {k^2 \over a^2} + M^2_{\rm eff}$.  The energy density in
these particles is then
\begin{equation} \label{energy}
\rho_\psi(t) = {2 \over a^3} \int{{{d^3k} \over {(2\pi)^2}} \Omega_k n_k(t)}
= {1 \over {a^3 \pi}} \int{ dk \, k^2 \Omega_k n_k(t)} \ .
\end{equation}
The normalization of the
solutions $X_k(t)$ is such that
$X_k(t \rightarrow 0^-) = N_+ e^{-i\Omega_k t}$ and $n_k(0)=0$.  Thus,
we find $N_+ = (2\Omega_k(\Omega_k - M_{\rm eff}))^{-1/2}$.  These are the
so-called positive frequency initial conditions.

A comment about the regularization of fermion VEVs.
In principle,  fermionic VEVs like $\langle T^{\mu\nu} \rangle$ require
regularization , which in the presence of background metrics and scalars
is rather non-trivial, see e.g. \cite{renorm} and references therein.
We will consider only the processes of particle creation, ignoring
vacuum polarization. The creation of particles, which corresponds to the 
imaginary part of the effective action, has no formal divergencies 
and does not require regularization. Assuming  the particle creation process 
dominates over vacuum polarization,
we will not consider the issues of regularization.

\section{\label{Noexp}  Fermionic Preheating without Expansion
    of the Universe}

It is convenient to begin the investigation of fermionic preheating
due to an oscillating scalar field
with a simplified setting neglecting the expansion of the universe.
This setting may have not only methodological advantages. Indeed,
    whenever the frequency of $\phi$ oscillations is much greater
than the rate of cosmic expansion $H$, it is sensible to neglect the time
dependence of the scale factor in solving the mode equation~(\ref{mode_exp}).
This can be the case, for example, when 
spontaneous symmetry breaking occurs rapidly leaving the $\phi$-field
oscillating about a new minimum where the effective mass $m_\phi$ happens
to be much larger than the Hubble parameter at the time of the transition.
One such example is hybrid inflation scenarios which end with TeV scale energy
densities.  Another example is the inflationary model
with the potential $\lambda \phi^4$. This theory possesses conformal
properties: at the stage of inflaton oscillations equations for the
fields by means of conformal transformations can be reduced to the
equations in Minkowskii space-time, see e.g. \cite{GK}.

In this  paper we will mostly use a chaotic inflationary model with
quadratic potential
 $V(\phi) = {1 \over 2} m^2_\phi \phi^2$.
  If we make the replacement $a=1$ in all the formulas of
Section 2, all the effects of expansion will be removed.
Background oscillations take the form of harmonic oscillations
 $\phi(t) = \phi_0 f(t)$ with $f(t)=\cos(m_\phi t)$ and
$\phi_0$ the time independent amplitude.  It is convenient to
define a new, dimensionless time variable $\tau \equiv m_\phi t$.
With this change of variables, the mode equation~(\ref{mode_exp})
may be written
\begin{equation} \label{mode}
X''_k + \left[ \kappa^2 + (\stackrel{\sim}{m} + \sqrt{q} f)^2 -i\sqrt{q}f'
\right] X_k =0 \ ,
\end{equation}
where we have introduced the dimensionless momentum $\kappa \equiv
{{k} \over {m_\phi}}$, the dimensionless fermion mass $\stackrel{\sim}{m}
\equiv {m_\psi \over m_\phi}$, and the resonance parameter
$q \equiv {{h^2 \phi^2_0} \over {m^2_\phi}}$.  These three parameters
completely determine the strength of the effect.  In fact,
this form of the mode equation is valid not only for harmonic
background oscillations, but
for generic $\phi$ oscillations in
a general potential.  For this, we identify $m_\phi$ with the frequency of
oscillation, $\phi_o$ with its amplitude, and $f(\tau)$ with the periodic
background oscillations normalized to unit amplitude.  Note that the
frequency will be amplitude-dependent for a general, non-quadratic
potential.

\subsection{\label{exc} Parametric Excitation of Fermions }

\begin{figure}[tb]
\begin{minipage}[t]{8.0cm} %{7.2cm}
   \centering \leavevmode \epsfxsize=8.0cm    %7.2cm
   \epsfbox{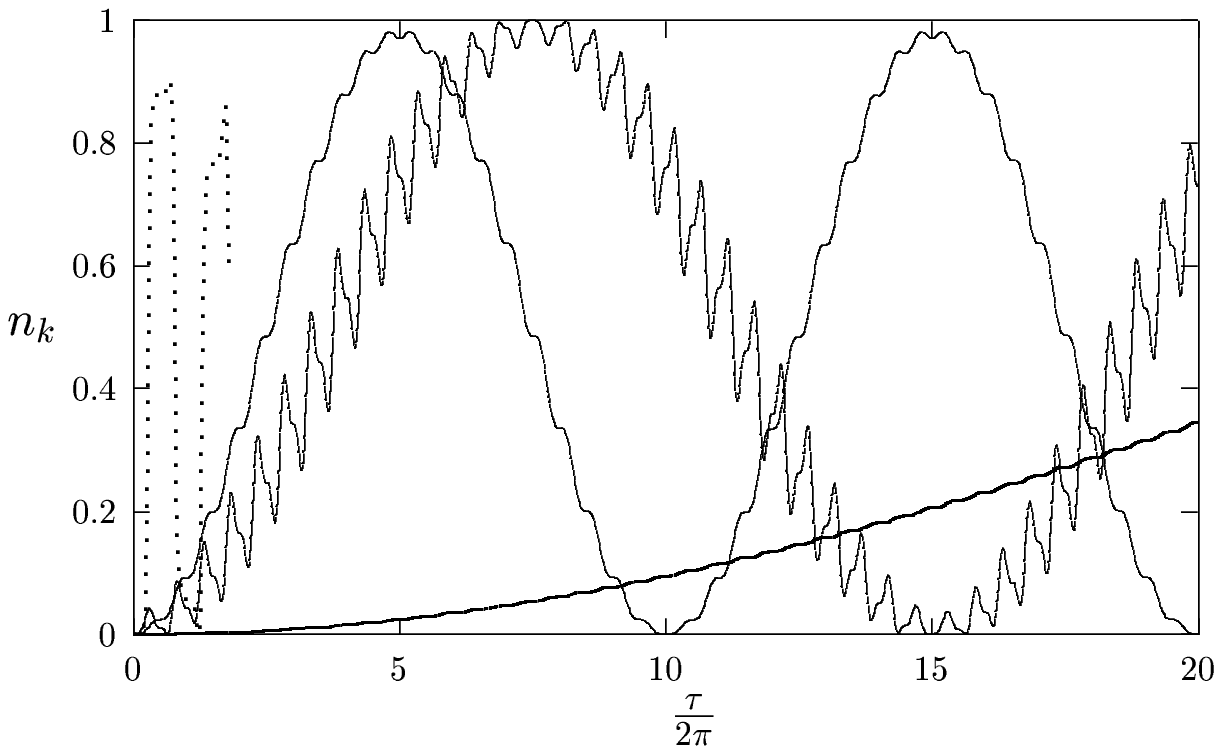}\\

   \caption[fig1u]{\label{fig:osc} {\em 
The occupation number $n_k$ of fermions in $m_\phi^2 \phi^2$-theory
as function of time ${\tau \over {2\pi}}$
for a range of $(q; \kappa)$: 
dotted curve for $(1.0; 0.05)$, bold curve for  $ (10^{-4}; 0.5)$,
smooth sine-like curve for $ (10^{-2}; 0.5)$, wiggly
  sine-like curve for $ (1; 1.3)$.
\hspace*{\fill}}}
\end{minipage}
\hfill
\begin{minipage}[t]{8.0cm}
   \centering \leavevmode \epsfxsize=8.0cm
   \epsfbox{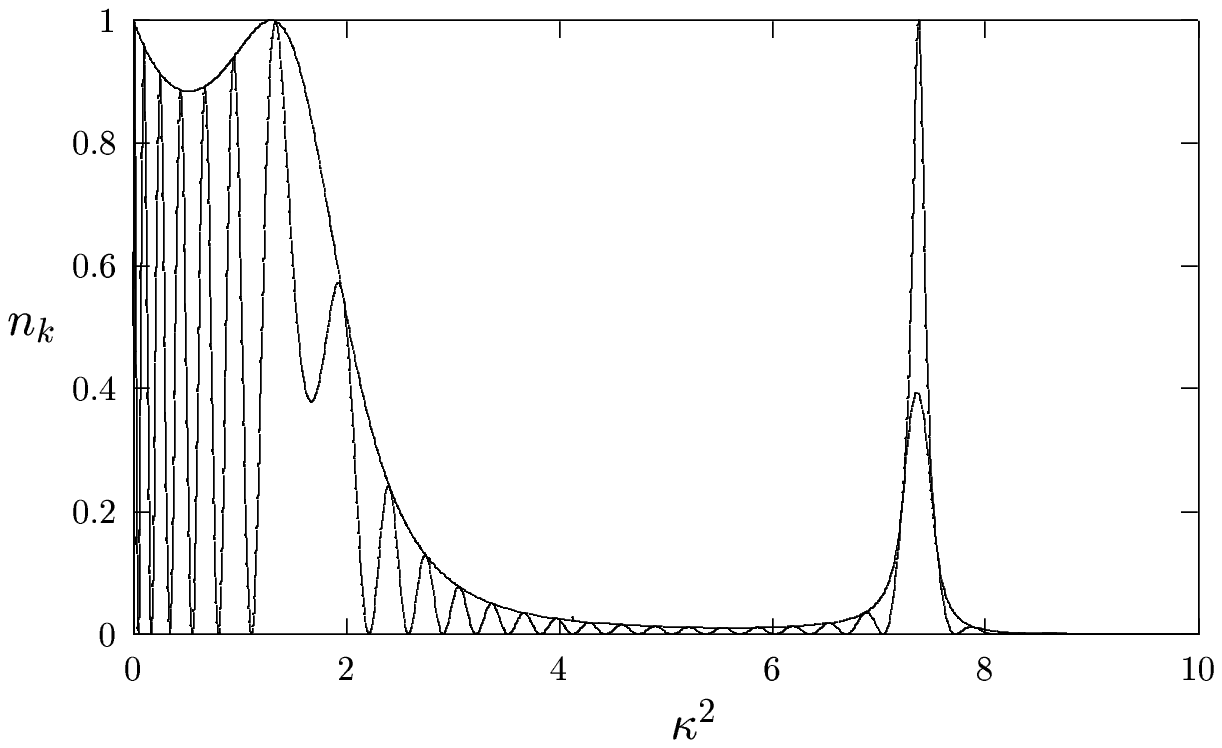}\\
   \caption[fig1u]{\label{fig:fill} {\em 
The comoving occupation number of fermions in $m_\phi^2 \phi^2$-inflation 
as a function of $\kappa^2$ after $10$ inflaton 
oscillations for resonance parameter $q = 10$.  Also shown is the
envelope function $F_k$ defined by equation~(\ref{15}).
  \hspace*{\fill}}}
\end{minipage}
\end{figure}

If individual inflatons at rest are decaying into light fermions in
a process $\phi \to \bar \psi\psi$, perturbative calculations
give the rate of decay $\Gamma_{ \phi \to \psi \psi }\simeq {{ h^2
m}\over {8\pi}}$ and the spectrum of created fermions is 
sharply peaked around $m/2$ with the width
 $\Gamma_{ \phi \to \psi \psi }^{-1}$.
In Figs.~\ref{fig:osc}, \ref{fig:fill}, however, we plot the time-dependence and
 spectrum of occupation number
for fermions created from the coherently oscillating inflaton field,
as follows from numerical solution of Eq. (\ref{mode}) and (\ref{nk}).
% Figures of oscillations and fill.
%---------------------------------------------
%---------------------------------------------
% End of Figures
%---------------------------------------------
The spectrum and time evolution  are drastically different  from what is
expected from the perturbative calculations. We therefore can
talk about a specific phenomena, the parametric excitation of fermions
interacting with coherent background oscillations.

% Figures of lphi4 mu_k and tdep.
%---------------------------------------------
\begin{figure}[tb]
\begin{minipage}[t]{8.0cm} %{7.2cm}
   \centering \leavevmode \epsfxsize=8.0cm    %7.2cm
   \epsfbox{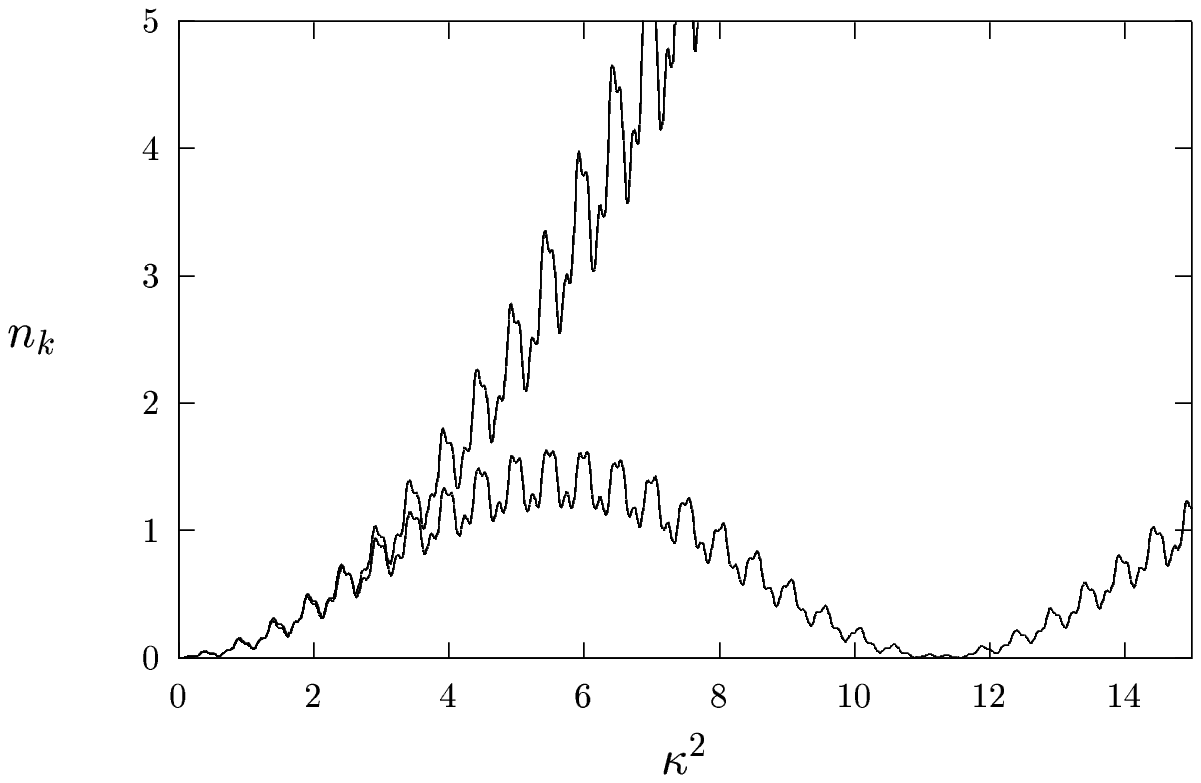}\\

   \caption[fig1u]{\label{fig:bosetdep} {\em 
Time dependence of the occupation number in $\lambda \phi^4$-theory
for $q=3$ for two different modes:  one is 
just inside the resonance band which grows like $\sinh(\mu_k t) \sim e^{\mu_k \tau}$ and
another  outside which oscillates like $\sin(\mu_k t)$.
\hspace*{\fill}}}
\end{minipage}
\hfill
\begin{minipage}[t]{8.0cm}
   \centering \leavevmode \epsfxsize=8.0cm
   \epsfbox{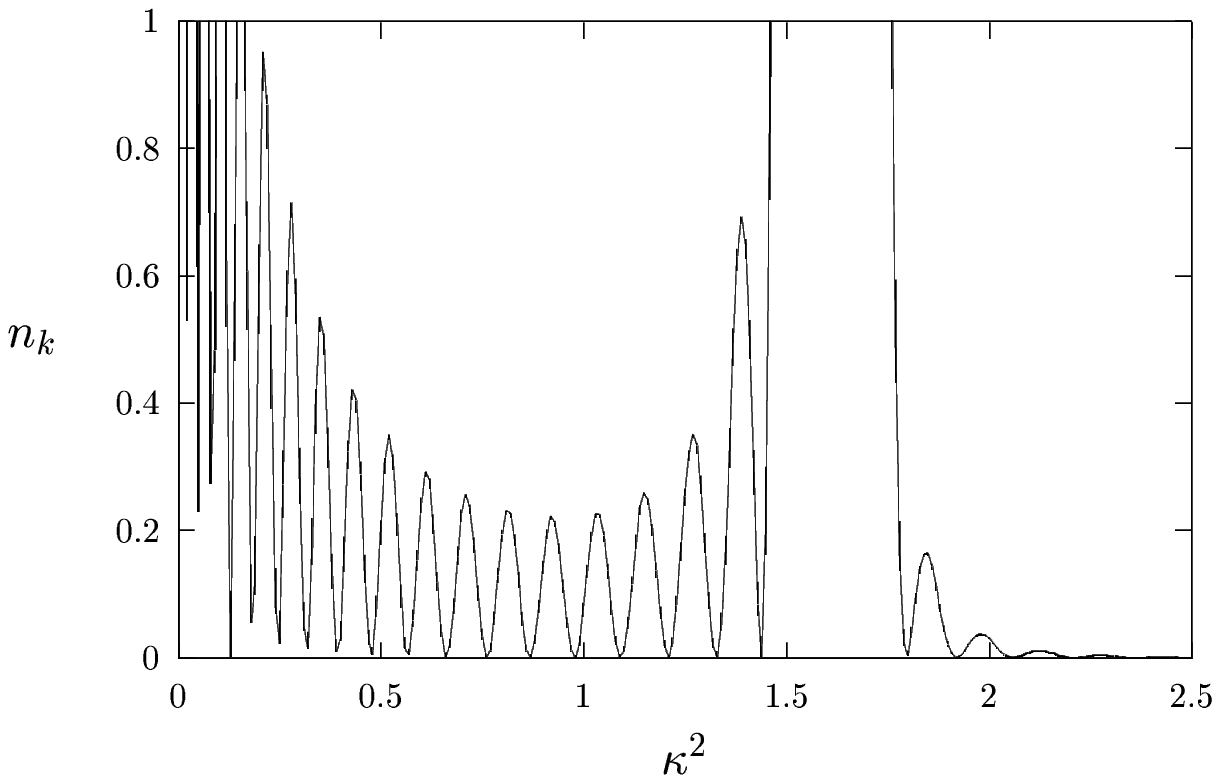}\\
   \caption[fig1u]{\label{fig:boseband} {\em 
Bosonic occupation number after 10 background oscillations
 for $\lambda \phi^4$-theory when
$q={g^2 \over \lambda}=3$.  A stripe with sharp edges
corresponds to the $e^{\mu_k \tau}$ instability.
  \hspace*{\fill}}}
\end{minipage}
\end{figure}
%---------------------------------------------
% End of Figures
%---------------------------------------------
It is instructive to compare the spectrum and evolution of fermionic
occupation number with those of bosonic occupation number.
The mode function of a quantum bose scalar field $\hat \chi$ coupling 
as $g^2\chi^2 \phi^2$
with oscillating
inflaton obeys the bosonic oscillator-like equation
\begin{equation} \label{modeb}
X''_{b,k} + \left[ \kappa^2+{\stackrel{\sim}{m_b}}^2 + q_b f^2
\right] X_{b,k} =0 \ ,
\end{equation}
where $q_b={{g^2 \phi_0}\over m^2}$,
${\stackrel{\sim}{m_b}}={m_{\chi}\over m}$.
In Figs.~\ref{fig:boseband} and \ref{fig:bosetdep}
we plot  the spectrum and time evolution
of bosonic  occupation number calculated from
 bosonic oscillator-like equations 
(\ref{modeb}). We use the familiar model of bosonic 
resonance due to the self-interaction in
$\lambda \phi^4$ inflation, which corresponds to $g^2=3\lambda$,
$q_b=3$, $ m_b=0$, $f(\tau)$ is given by oscillations in
$\lambda \phi^4$ theory. 

In the bosonic case,
there are distinct resonance bands in
 which bosonic modes are exponentially unstable,
$n_k^b(t) \sim e^{{\mu_k}\tau}$. Particle
creation takes place outside of the resonant bands as well,
although there the occupation number is bounded and oscillates periodically.
In the fermionic case  $n_k \leq 1$ is always bounded by Pauli
blocking and oscillating periodically with time.
The occupation  number in both cases is changing in time. It is
therefore convenient to introduce an envelope function $F_k$
of the particle spectrum \cite{GK},
which corresponds to $n_k$ averaged over short-time intervals (order of the 
background oscillation period).
A bosonic envelope function cannot be defined for the resonant bands
(where it is $e^{{\mu_k}\tau}$.)  This zone corresponds to the gap between almost
vertical  lines
in Fig.~\ref{fig:boseband}.
The fermionic envelope function $F_k$, on the contrary, can be defined everywhere.
Although it is always bounded by unity, their structure is reminiscent of the
resonant-band structure:
for some range of $k$, the `resonant' band,
$n_k $ is close to unity, while in other ranges, the stable bands, it is significantly
smaller if not zero. In \cite{hybrid} different levels of $F_k$ were
plotted on the parameter plane $(\kappa, q)$, revealing a structure
which reminiscent of the stability/instability chart of the
bosonic parametric resonance.
One of the most important results is that
fermionic parametric excitations occurs very quickly,
within about ten(s) background oscillations.
Interestingly,  fermionic "resonant bands" are excited the last,
while non-resonant intervals fill first.

Comparison of bosonic and fermionic parametric  excitations
is useful to understand some features of bosonic preheating.
As we have seen, there is production of bosons outside of the resonance band.
If we take into account the expansion of the universe, in the most interesting case
of large $q_b$, the difference between resonant and non-resonant excitations
 of bosons  will be erased, and the regime of stochastic resonant production of
bosons will be settled down \cite{KLS97}.

\subsection{\label{analytic} Some Generic Analytic Results}

Dynamics of the Fermi field coupling to the background  homogeneous
scalar can be revealed with the second-order oscillator-like equation
(\ref{mode})
for the mode function $X_k(t)$. For periodic background oscillations
$f(t)=f(t+T)$, $T$ is a period,  some generic analytic
results were derived a long time ago \cite{em} in the context of
particle creation in a periodic external electromagnetic field.
In particular, the occupation number of created particles
at instances $t=N_sT$, i.e. exactly after $N_s$ background oscillations,
  is given by expression \cite{em}
\begin{equation}
n_k(N_sT)  = {k^2 \over {2\Omega_k^2}}{{ \sin^2 N_s d_k} \over { \sin^2  d_k}}
 \left({\it Im} X_k^{(1)}(T)\right)^2 \ ,
\label{occ}
\end{equation}
where
$\cos d_k={\it Re} X_k^{(1)}(T)$.
To derive this result, one introduces two
fundamental solutions of Eq.~(\ref{mode}),
$ X_k^{(1)}(t)$ and $ X_k^{(2)}(t)$;
such that initially
$ X_k^{(1)}(0)=1$, $ \dot X_k^{(1)}(0)=0$ and
 $ X_k^{(2)}(0)=0$, $ \dot X_k^{(2)}(0)=1$.
Expression~(\ref{occ}) involves only
 the value of the first
fundamental solution $X_k^{(1)}(T)$ exactly after the first
oscillation.
It says that the occupation number of created particles
 after $N_s$ background
oscillations is modulated with a certain frequency $\nu_k$
(which, as we will see, does not coincide  with $d_k$).
 However, practical application of the 
generic formula (\ref{occ}) is rather limited, because
it does not address the full time evolution of $n_k(t)$, and
cannot strictly determine a period of modulation $\pi /\nu_k$.

To get an idea of how the 
 occupation number of created particles $ n_k(t)$  evolves with  time,
again let us look at  Fig.~\ref{fig:osc} and further at  Fig.~\ref{fig:parabolic}
 for
different values of the parameters $\kappa$ and  $q$.
For small and moderate $q$ (but not too small $\kappa$)
the occupation number exhibits
high frequency (period $< {T \over 2}$)
oscillations which are modulated by a long period behavior.
For large $q $ number of fermions jumps in a step-like manner
at instances when effective mass of the fermi field crosses zero,
superposed by very high frequency oscillations around almost
constant values,
as depicted in Fig.~\ref{fig:parabolic}.
These  jumps are modulated with a frequency  $\nu_k/2$:
steps in the first half of the cycle up are accumulated until
$n_k(t)$ reaches its maximum $F_k$, and then steps down  to zero
in the second half of the cycle.

However, this picture of high frequency features superimposed over
long-period modulation is not universal.
For small $\kappa$, or for one of the most interesting cases of
$q \gg 1$ and moderate $\kappa$,
the occupation 
number of fermions jumps between zero and one
within time interval much shorter than period of background
oscillations,
 as depicted by the dotted curve in
Fig.~\ref{fig:osc}.
  There are interesting situations when fermions
are created in an instant (single kick) process, where
formula (\ref{occ}) is not applicable. 
Therefore, for different ranges of parameters we will shall develop
different approaches.

\subsection{\label{moder} Semi-Analytic Theory for Averaged Occupation Number}

Numerical curves for the mode functions suggests splitting of the
time evolution into higher frequency features with the time-scale
comparable or less than the period of background oscillations $\leq T$,
and low-frequency modulations with the period $\pi/\nu_k$ greater than $T$.
In this case we  can utilize generic result (\ref{occ}).
However, it is convenient to use not the values  $n_k(N_sT)$, but rather
 the smoothed  occupation number $\bar n_k(t)$ which is $n_k(t)$
averaged over  high frequency
oscillations,
$\bar n_k(\tau)= { 1 \over T} \int_{\tau}^{(\tau+T)} d \tau  n_k(\tau)$.
Then we can write the smoothed occupation number of fermions
in a factorized form
\begin{equation}
\bar n_k(\tau) = F_k \sin^2  \nu_k \tau \ ,
\label{average}
\end{equation}
where we introduce an  envelope function $F_k$.
The  average occupation number
of fermions evolves {\em periodically} with time.
The spectrum of $\bar n_k$ can be characterized by the 
envelope function $F_k$ 
and the period  of modulation 
${\pi \over \nu_k}$ which depends also on the parameter $q$.

Now we will utilize the result (\ref{occ}). 
We found that 
 the envelope function can be extrapolated by the
factor in the front of ${\sin^2 N_sd_k}$ in (\ref{occ}) and given by the 
 expression
\begin{equation}
F_k={1 \over {\sin^2 \nu_k T}} \,
{\kappa^2 \over {2 \Omega_k^2}} \,  \left({\it Im} X_k^{(1)}(T)\right)^2  \ .
\label{15}
\end{equation}
Next is to determine the frequency $\nu_k$ of the $\bar n_k$ modulations.
The value $d_k$ defined after  (\ref{occ}) cannot be the right answer,
because it would incorrectly predict that the 
peaks of $F_k$ are filled up first, while actually they are filled 
last. We tried the combination $\nu_k=\pi/2-d_k$, because
it corresponds to correct order of saturation of  $F_k$, and it occurs to
 work well. Therefore, the  modulation frequency $\nu_k$ is given 
by the relation $\cos \nu_k T = -{\it Re} X_k^{(1)}(T)$.
Thus, to find $F_k$ and $\nu_k$,
one need only calculate the complex value
$ X_k^{(1)}(T)$  after a single  background oscillation,
instead of performing a full numerical
integration of Eq.~(\ref{mode}).

\begin{figure}[tb]
\begin{minipage}[t]{8.0cm} %{7.2cm}
   \centering \leavevmode \epsfxsize=8.0cm    %7.2cm
   \epsfbox{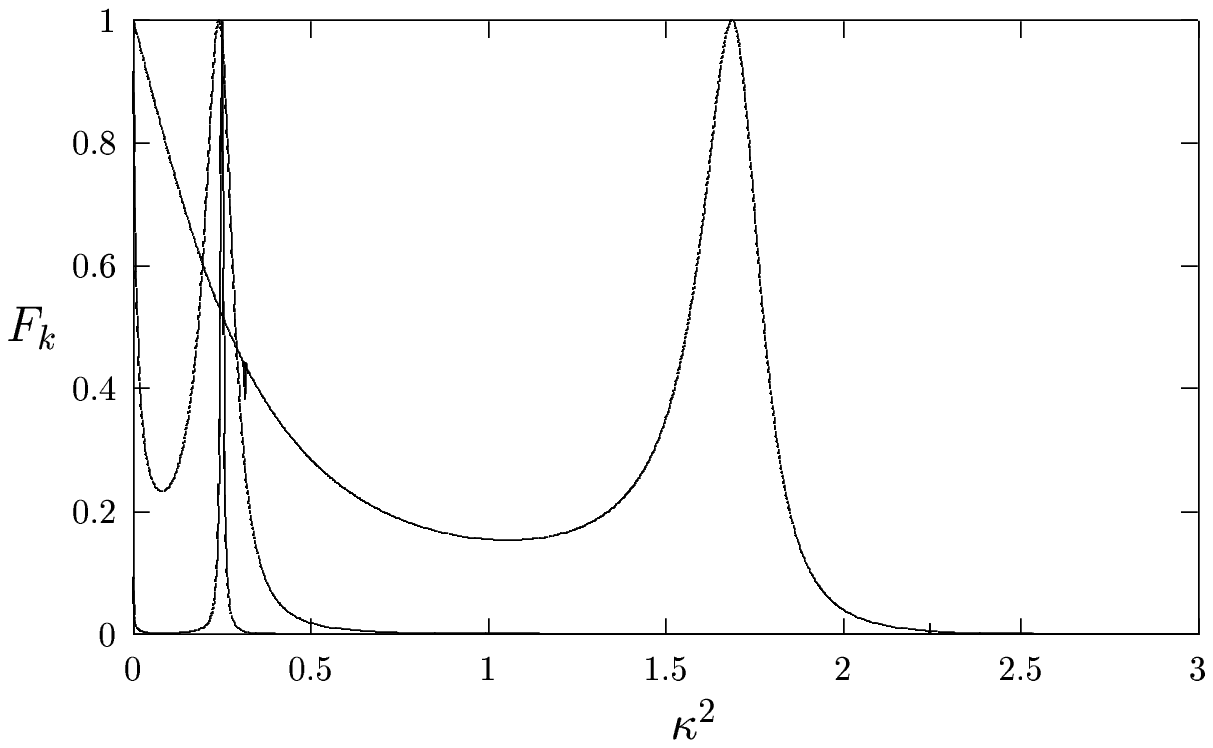}\\

   \caption[fig1u]{\label{fig:envel} {\em 
The envelope function $F_k$ for the amplitude of occupation number
oscillations in $m_\phi^2 \phi^2$-theory.  Values for
$q =  10^{-4}, 10^{-2}$ and $1$ are shown, from narrowest to broadest, respectively.
\hspace*{\fill}}}
\end{minipage}
\hfill
\begin{minipage}[t]{8.0cm}
   \centering \leavevmode \epsfxsize=8.0cm
   \epsfbox{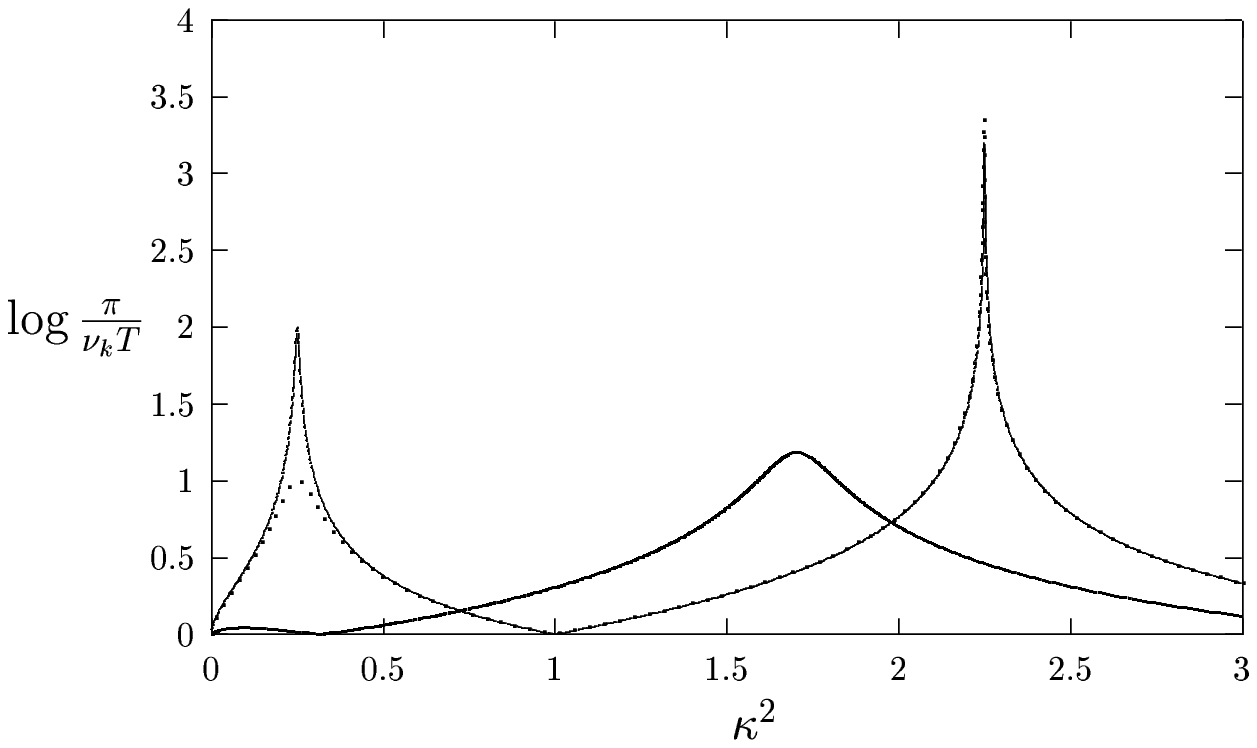}\\
   \caption[fig1u]{\label{fig:period} {\em 
The log of the modulation period in  $m_\phi^2 \phi^2$-theory,
where $q =1$ corresponds to the middle curve, 
 $q =10^{-2}$ corresponds to the dotted curve, 
and  $q =  10^{-4}$ corresponds to the remaining.
  \hspace*{\fill}}}
\end{minipage}
\end{figure}

We calculated $ X_k^{(1)}(T)$ numerically for ${ 1\over 2}m^2\phi^2$
 background model  and
constructed the envelope  function $F_k$ 
plotted in  Fig.~\ref{fig:envel} (similar graph was plotted in \cite{GK}
for ${ 1\over 4}\lambda\phi^4$ model). 
In  Fig.~\ref{fig:fill} we show,
using  (\ref{15}), how the 
fermionic resonance bands are filled after
$10$ background oscillations. 
% Figures of envelope and period.
%---------------------------------------------
%---------------------------------------------
% End of Figures
%----------
The function $\nu_k$ gives us the time scale
for fermion excitation.
In Fig.~\ref{fig:period} we plot the period
of modulation $\pi \over \nu_k$ as a function of $k$.
This function is peaked where $F_k$ is peaked, i.e.
the peaks of the resonance curve are the last to fill.

\subsection{\label{broa} Method of Successive Scatterings for Fermions}

It turns out that for large values of the parameter 
 $q$ we can significantly advanced in calculations of $n_k(t)$
beyond 
 the results of Section (\ref{moder}). One can expect that 
 $q$ may be much greater that one. Indeed, $q={{h^2 \phi_0^2} \over m^2}$.
In the context of chaotic inflation with the potential $V(\phi)={1 \over 2} m^2 \phi^2$
it follows from the theory of cosmological perturbations that
${\phi_0^2 \over m^2} \simeq 10^{12}$. On the other hand one can admit that Yukawa
 coupling can be $h \gg 10^{-6}$, which provides $q \gg 1$.

We will  generalize for fermions 
the method of parabolic scatterings
introduced for the bosonic resonance in reference~\cite{KLS97}.  The method
is based on the observation that, for $q \gg 1$, change of the particle number
occurs only during a short time interval $\tau_*$ near the zeros of the
effective mass of the particles.  This occurs because equation~(\ref{nk})
 for the number of
particles in terms of the mode functions is an adiabatic
invariant of the mode equation. 

\begin{figure}[tb]
%\begin{minipage}[t]{8.0cm} %{7.2cm}
   \centering \leavevmode \epsfxsize=8.0cm    %7.2cm
   \epsfbox{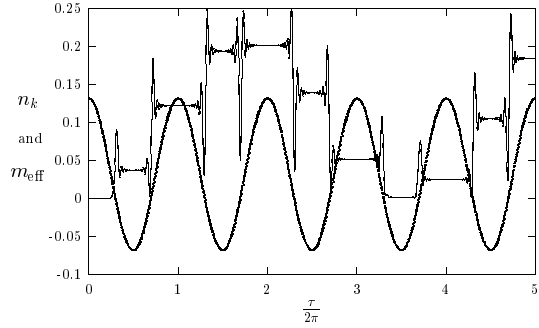}\\

   \caption[fig1u]{\label{fig:parabolic} {\em 
$n_k$ and 
$M_{\rm eff}(\tau)=
h\phi(\tau)+m_{\psi}$
 for $q=100$ and $\kappa=\sqrt{q} \approx 31.6$.
\hspace*{\fill}}}
%\end{minipage}
\hfill
\end{figure}

  For large $q$, the adiabaticity
condition $\Omega_k' < \Omega^2_k$ is violated only near the times $\tau_*$
when the effective mass vanishes.  This leads to a step-like evolution in
the number of fermions.  This is illustrated in figure~(\ref{fig:parabolic})
which shows the evolution of $n_k(\tau)$ and $M_{\rm eff}(\tau)=
h\phi(\tau)+m_{\psi}$

% Figures of parabolic.
%---------------------------------------------
%---------------------------------------------
% End of Figures
%---------------------------------------------

Near the times $\tau_*$, we may approximate the
effective mass in equation~(\ref{mode}) by
$(\stackrel{\sim}{m} + \sqrt{q} f) =
{1 \over m_{\phi}}\left(m_{\psi}+h\phi(\tau)\right) \approx {h \over m_{\phi}}
\phi(\tau_*)'(\tau - \tau_*)+O((\tau - \tau_*)^2)$.
We may also write this as
\begin{equation}\label{der}
(\stackrel{\sim}{m} + \sqrt{q} f) \approx
 \sqrt{q} f'(\tau_*)(\tau - \tau_*)= 
 \pm(q - \stackrel{\sim}{m}^2)(\tau - \tau_*) \ ,
\end{equation}
 where
the sign on the left hand side depends on whether $f(\tau)$ is
increasing ($+$) or decreasing ($-$) at $\tau_*$. 
Thus, in the neighborhood of $\tau_*$, the mode equation becomes
\begin{equation} \label{parabolic}
X_k'' + \left[ \kappa^2 + (q - \stackrel{\sim}{m}^2)^2(\tau - \tau_*)^2
-i \, {\rm sgn}(f')\sqrt{q - \stackrel{\sim}{m}^2} \, \right] X_k = 0 \ ,
\end{equation}
which is a Schroedinger equation for scattering off a negative
parabolic potential centered at $\tau_*$.

	The method of parabolic scattering
uses the exact solution of equation~(\ref{parabolic}) to provide a
connection formula between the adiabatic approximations of the full
mode equation on either side of $\tau_*$.  Suppose the $j$-th 
zero of $M_{\rm eff}$ occurs at time $\tau_j$.  For times
between $\tau_{j-1} < \tau < \tau_{j}$,  the general solution of the mode
equation~(\ref{mode}) takes the adiabatic (or semi-classical) form
\begin{equation}
X_k^{j}(\tau) ={\alpha_k^{j}} N_+ \,
 e^{-i\int_0^\tau d \tau \, \Omega_k(\tau) } +
{\beta_k^{j}} N_- \, e^{+i\int_0^\tau d\tau \, \Omega_k(\tau) } \ ,
\label{WKB1}
\end{equation}
where the coefficients $\alpha_k^{j}$ and $\beta_k^{j}$
are constant for $\tau_{j-1} < \tau < \tau_j$. 
After the scattering at $\tau_j$, $X_k(t)$, within the interval
 $\tau_j < \tau < \tau_{j+1}$, again has the adiabatic form
\begin{equation}
X_k^{j+1}(\tau) ={\alpha_k^{j+1}} N_+ \,
 e^{-i\int_0^\tau d \tau \, \Omega_k(\tau) } +
{\beta_k^{j+1}} N_- \, e^{+i\int_0^\tau d\tau \, \Omega_k(\tau) } \ ,
\label{WKB2}
\end{equation}
with new coefficients $\alpha_k^{j+1}$ and $\beta_k^{j+1}$
that are constant for $\tau_{j} < \tau < \tau_{j+1}$.
At $\tau=0$, our vacuum positive frequency condition requires $\alpha^1_k = 1$
and $\beta^1_k =0$.  Particle creation occurs when, after scattering at the
times $\tau_j$, the initial positive
frequency wave acquires a negative frequency part.  The number density
of produced particles with momentum $k$ is $n_k^j = \vert \beta_k^j
\vert^2$ for times $\tau_{j} < \tau < \tau_{j+1}$.  Furthermore,
normalization requires
$\vert \alpha_k^j \vert^2 + \vert \beta_k^j \vert^2 = 1$ for all $j$.

  The important observation is that the outgoing amplitudes
 ($\alpha_k^{j+1} $, $ \beta_k^{j+1}$) can be expressed through
 the incoming amplitudes
($\alpha _k^{j}$, $ \beta_k^{j}$) by means of the reflection
 $R_k$ and transmission $D_k$
amplitudes for scattering at $t_j$:
\begin{equation}\label{matrix}
\pmatrix{\alpha_k^{j+1} e^{-i \theta_k^{j}}
 \cr \beta_k^{j+1} e^{+i \theta_k^{j}}
 \cr } =
\pmatrix{ {1 \over D_k} & {R^*_k \over D^*_k} \cr
 {R_k \over D_k} & {1 \over D^*_k} \cr}
\pmatrix{\alpha_k^{j}e^{-i \theta_k^{j}} \cr \beta_k^{j}
e^{+i\theta_k^{j}} \cr}
 \ .
\end{equation}
Here $\theta_k^{j}=\int\limits_0^{\tau_j} d\tau \, \Omega(\tau)$ is the phase
accumulated by the moment ${\tau_j}$.

	If we let $x=(q - \stackrel{\sim}{m}^2)^{1/4}\tau$,
equation~(\ref{parabolic}) may be written
\begin{equation}
{d^2 X_k \over d{x}^2} + \left[ \Lambda_k^2 -i(-1)^j +
 x^2 \right] X_k =0 \ ,
\label{parabolic2}
\end{equation}
where the parameter
\begin{equation}
\Lambda_k^2 \equiv {\kappa^2 \over 
|hf'(\tau_*)|} = {\kappa^2 \over 
\sqrt{q - \stackrel{\sim}{m}^2}} \ . \nonumber
\end{equation}
A general analytic solution of equation~(\ref{parabolic2})
is a linear combination of the parabolic cylinder functions
 $ W \left(-{{(\Lambda_k^2 -i(-1)^j)} \over 2} ; \pm\sqrt{2} x \right)$.
The reflection $R_k$ and transmission $D_k$
amplitudes for  scattering on the parabolic potential
can be found from the asymptotic forms of these analytic solutions:
\begin{equation}\label{R}
R_k=- { i e^{ i \varphi_k} \over {\sqrt{1-e^{\pi \Lambda_k^2} }} } \ ,
D_k= { e^{-i \varphi_k} \over {\sqrt{1-e^{-\pi \Lambda_k^2}}}} \ ,
\end{equation}
where the angle $\varphi_k$ is
$\varphi_k= \arg
 \Gamma \left({1+i\Lambda_k^2 \over 2}\right)+{\Lambda_k^2\over 2}
\left(1+ \ln{2\over \Lambda_k^2}\right)$.
Note that the angle $\varphi$ depends on the momentum $k$.
Substituting (\ref{R})  into (\ref{matrix}), we
can determine the change in induced in the $\alpha_k$ and $\beta_k$
coefficients by a single parabolic scattering 
in terms of the parameters of the parabolic potential
and the phase $\theta_k^{j}$ only.
Specifically, we find 
\begin{eqnarray}\label{matrix1}
&&\pmatrix{\alpha_k^{j+1} \cr \beta_k^{j+1} \cr } =% \\
% &&
\pmatrix{ \sqrt{1-e^{-\pi \Lambda_k^2}} e^{i\varphi_k} &
 -(-1)^j e^{-{{\pi \over 2} \Lambda_k^2} +2i\theta_k^{j}} \cr
 (-1)^j e^{-{{\pi \over 2} \Lambda_k^2} -2i\theta_k^{j}}
 & \sqrt{1-e^{-\pi \Lambda_k^2}} e^{-i\varphi_k} \cr}
\pmatrix{\alpha_k^{j} \cr \beta_k^{j} \cr}
 \ . 
\end{eqnarray}
It is now a simple matter to find the change in particle number
after one scattering.  From the normalization of $\alpha_k$ and $\beta_k$
we have the relation
\begin{equation}
\begin{array}{rclcccc} \left(\begin{array}{cc}
{\alpha^*}_k^{j+1} & {\beta^*}_k^{j+1}
\end{array}\right) &
\left(\begin{array}{cc}
 1 & 0 \\
 0 & -1 
\end{array}\right) &
\left(\begin{array}{c}
\alpha_k^{j+1} \\ \beta_k^{j+1}
\end{array}\right) &
= & {\vert \alpha_k \vert^2 - \vert \beta_k \vert^2} & = &
{1 - 2n_k^{j+1}} 
\end{array} \ ,
\end{equation}
which, when applied to equation~(\ref{matrix1}) gives the desired
result
\begin{eqnarray}
n^{j+1}_k &=& e^{-\pi \Lambda_k^2} + \left( 1 -  2 e^{-\pi \Lambda_k^2}
\right) n^{j}_k\nonumber \\ &-&
2 (-1)^j e^{-{\pi \over 2} \Lambda_k^2}\sqrt{1 - e^{-\pi \Lambda_k^2}}
\sqrt{ n^{j}_k (1 - n^{j}_k)} \sin \theta_{tot}^{j} \ ,
\label{nkfermi}
\end{eqnarray}
where the phase $\theta_{tot}^{j} = 2\theta_k^{j}-\varphi_k + \arg
\beta_k^{j}-\arg \alpha_k^{j} $.
Let us discuss this formula, which is the main result of our paper.
If fermions are light, $m_{\psi}=0$, their occupations number is changing with time
only when background field crosses zeros. Without expansion of the universe,
the phases $\theta$  accumulated between successive zeros are equal.
In this case one can try to proceed to find the solution of the matrix equation
(\ref{matrix}), as it was done for bosons \cite{KLS97}. However, for bosons
that was needed to find the stability/instability bands, which is not so interesting
for fermions.
In the case of massive fermions time intervals between successive zeros of
 $M_{eff}$
are not equal, and problem of finding matrix solution of eq. (\ref{matrix})
became even  more complicated.
However, in the most interesting case of expanding universe the phases
between successive zeros of $M_{eff}$ became random
(see \cite{KLS97} for details). Then we can just put $\theta_{tot}$ as a random
phase and use formula (\ref{nkfermi}) as it is.

\section{\label{exp} Parametric Excitation of Fermions with Expansion of 
    the Universe}

To address the problem of fermionic preheating after $m^2\phi^2$ chaotic
inflation, we must now deal with the full mode equation~(\ref{mode_exp}),
which no longer has a periodic time dependence in the complex frequency.
Nevertheless, it is still convenient to work with
the form~(\ref{mode}) in the time variable $\tau=m_\phi t$ where the
parameters $q$ and $\kappa^2$ are now understood to be time dependent:
\begin{equation} \label{mode_expan}
{X}_k'' + \left[ {\kappa^2 \over a^2}
+ M^2_{\rm eff} -i{{ {(a \, M_{\rm eff})}}' \over a}
+ \Delta(a) \right] X_k = 0 \ ,
\end{equation}
{\bf Check}
where
$'$ stands for the derivative in respect with $\tau$, $\kappa$ is
a comoving momentum scaled in units of $m_\phi$.
There  is often used another from of the fermionic mode equation
written in terms of conformal time $\eta=\int dt/a$ and
 mode function $Y_k=a^{-3/2}\chi_k$
\begin{equation} \label{conf} 
\partial_{\eta}^2Y_k + \left[ {\kappa^2 }
+ M^2_{\rm eff} -i  \partial_{\eta} {{ {(a \, M_{\rm eff})}}}
 \right] Y_k = 0 \ .
\end{equation}
This form of equation is useful, for instance, in conformal theory
$V(\phi)=\lambda \phi^4$
when the problem can be reduced to the problem in Minkowski space-time
 \cite{GK}. In this case background field is oscillating periodically 
in respect with time $\eta$.
In the case of quadratic inflaton potential background field
is oscillating  with the constant period in terms of physical time $t$
 (or $\tau$).
We therefore will use mode equation (\ref{mode_expan}). 
The
parameter $q$ is  now understood to be time dependent.
Specifically, we have
$q(\tau) \equiv {{h^2 \Phi^2(\tau)} \over m^2_\phi}$, scaled
physical momentum
$p \equiv {1 \over a} {k \over m_\phi}$
Here, $\Phi(\tau)$ is the time dependent amplitude of inflaton oscillations.
As is well known~(c.f. \cite{KLS97}), oscillations of
$\phi$ in this model quickly approach the asymptotic solution
$\phi(\tau) = \Phi(\tau) \, \cos(\tau)$,
$\Phi(\tau) \approx {\phi_0 \over {a^{3/2}}} \sim 
{M_{\rm pl} \over {\sqrt{3\pi} (1+\tau)}}$ 
where time is measured from the start of inflaton oscillations, 
$\phi_0 \approx {M_{\rm pl} \over {\sqrt{3\pi}}}$, and the scale factor
averaged over several oscillations behaves as in a matter dominated universe:
$a(\tau) = (1+\tau)^{2/3}$. 

\subsection{\label{broads} Stochastic Parametric Excitations of Fermions }

\begin{figure}[tb]
\begin{minipage}[t]{8.0cm} %{7.2cm}
   \centering \leavevmode \epsfxsize=8.0cm    %7.2cm
   \epsfbox{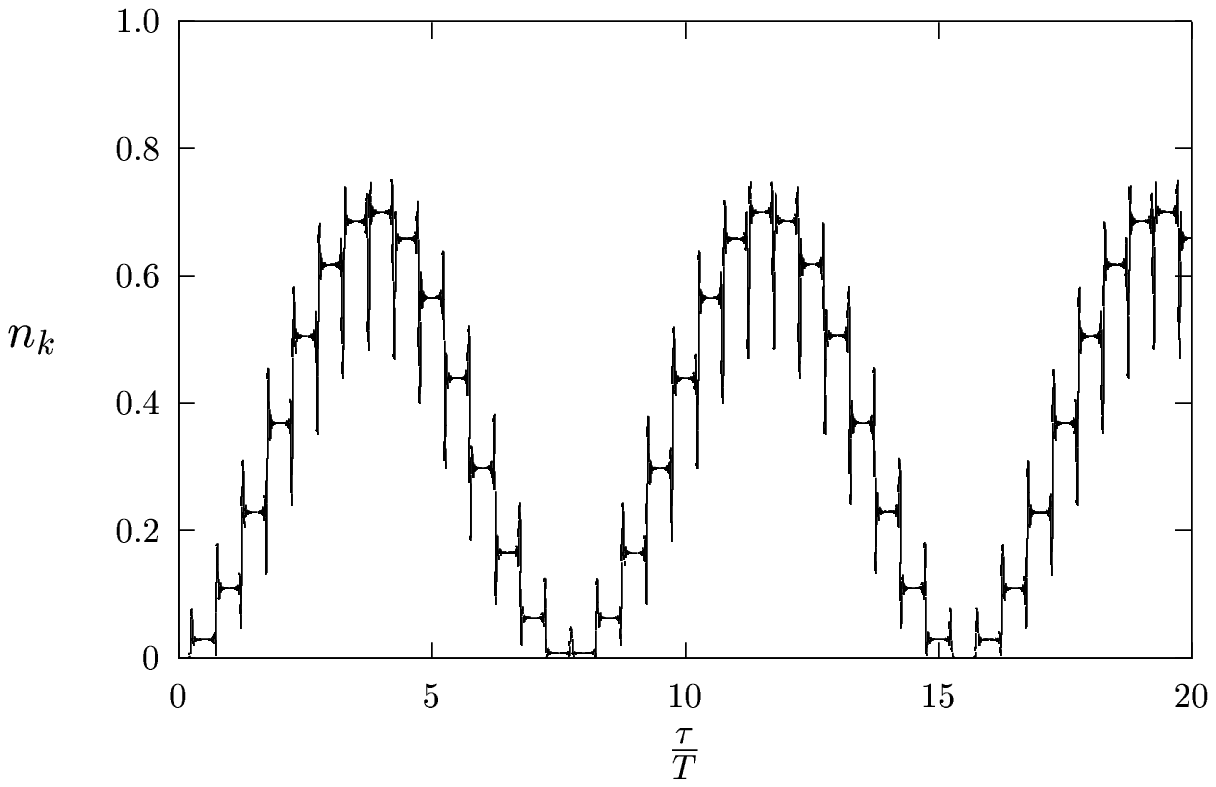}\\

   \caption[fig1u]{\label{fig:nostoch} {\em 
The occupation number $n_k$ of fermions in $m_\phi^2 \phi^2$-theory
without expansion.  Here $q=10^6$ and the particular mode is
$\kappa^2 \approx \sqrt{q}$.
\hspace*{\fill}}}
\end{minipage}
\hfill
\begin{minipage}[t]{8.0cm}
   \centering \leavevmode \epsfxsize=8.0cm
   \epsfbox{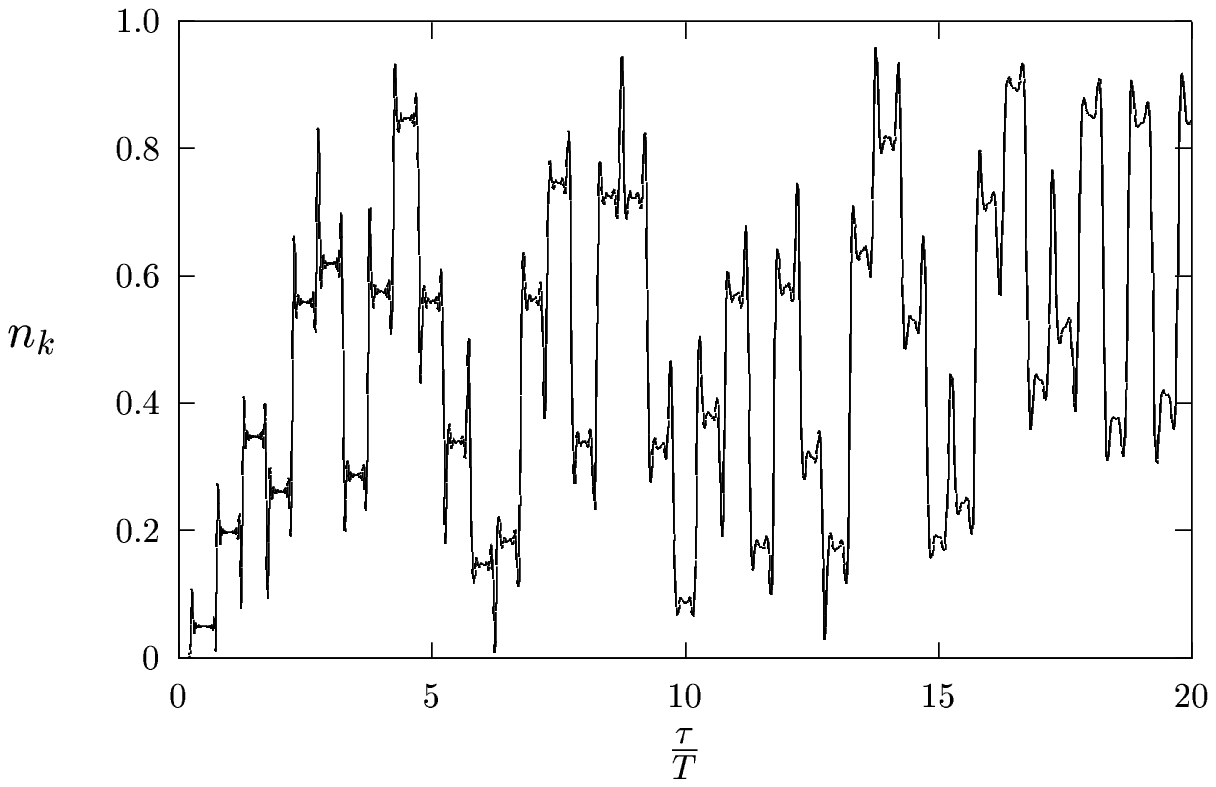}\\
   \caption[fig1u]{\label{fig:stoch} {\em 
The occupation number $n_k$ of fermions in $m_\phi^2 \phi^2$-theory
with expansion taken into account.  Here the initial resonance parameter
is $q_0=10^6$ and the mode is $\kappa^2_0 \approx \sqrt{q}$.

  \hspace*{\fill}}}
\end{minipage}
\end{figure}

        Let us first consider the case of light fermions with a large
initial resonance parameter: $h \phi_0 \gg m_\phi \gg m_\psi$.  
Fig.~(\ref{fig:nostoch}) shows a numerical solution of
equation~(\ref{mode})  in the absence of expansion
with a resonance paramter $q=10^6$.  As expected from for such a large $q$
parameter, we see a step-like change in occupation number with
periodic modulation.
% Figures of tdev for nostoch and stoch.
%---------------------------------------------
%---------------------------------------------
% End of Figures
%---------------------------------------------
In Fig.~(\ref{fig:stoch}), we show a numerical
solution to the  equation (\ref{mode_expan}) in the presence of expansion with
the {\em initial} parameters $q_0=10^6$. Here, the
{\em comoving} number density of particles is plotted.  We see that,
as for the non-expanding case, evolution between the zeros of the
effective mass (oe equally inflaton field),
 $M_{\rm eff} \approx \sqrt{q(\tau)} \cos(\tau)$, is
adiabatic.  We can thus apply the results of Section~III,D 
in particular the analytic formula~(\ref{nkfermi}), mutatis mutandis.
The Pauli principle is obviously still obeyed and
the typical step in particle number is still suppressed by a factor
$e^{-\pi \Lambda_k}$ which is now time dependent.  The most important
qualitative change is that the accumulated phase $\theta_{\rm tot}$ is
now uncorrelated between successive zeros of the effective mass 
(inflaton field).
This occurs
because the effective frequency $\Omega_k =\sqrt{p^2
+q^2(\tau)\cos^2\tau}$ is no longer periodic and the accumulated phase
$\theta=\int_{j}^{{j+1}} \Omega_k \approx {{2h\Phi(\tau)} \over m}
+ O(k^2)$
changes substantially
in magnitude within one inflaton oscillation,
$\delta \theta_k \simeq {{\sqrt q} \over {2N_s^2}}$,
after the $N_s$-th oscillation \cite{KLS97}.
  The result is that
the $\sin(\theta_{\rm tot})$ term in~(\ref{nkfermi}) becomes a 
random variable.  As is readily apparent in Fig.~(\ref{fig:stoch}), this
destroys the periodic modulation of $n_k$ and the parametric excitation of
fermions becomes stochastic, as anticipated in~\cite{GK}.

\begin{figure}[tb]
%\begin{minipage}[t]{8.0cm} %{7.2cm}
   \centering \leavevmode \epsfxsize=8.0cm    %7.2cm
   \epsfbox{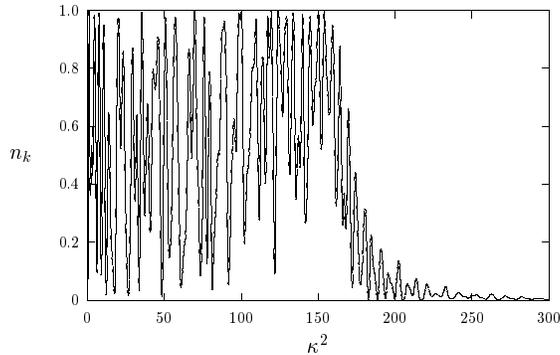}\\

   \caption[fig1u]{\label{fig:sphere} {\em 
The comoving occupation number of fermions in $m_\phi^2 \phi^2$-inflation 
after $50$ inflaton oscillations for initial resonance parameter $q_o = 10^3$.
Expansion destroys the details of the resonance band and
leads to a fermi-sphere of width $q^{1/4}$ which grows like $a^{1/4}$
while $q(\tau) > 1$.
\hspace*{\fill}}}
\end{figure}

       Once the periodic modulation of $n_k$ is destroyed, the 
construction of Section~III,C is no longer valid: the occupation
number cannot be characterized by an amplitude and period.  In fact, the
spectrum of created particles is even simpler.  Stochastic excitation
allows a given comoving mode $\kappa$ to obtain any amplitude in the
range $0 \le n_k \le 1$ if there are a sufficient number of parabolic
scatterings, $N_s$. This gives us  the picture of stochastically filling
a (Fermi) sphere in the momentum space. 
Numerical calculations confirm this picture, see Fig.~\ref{fig:sphere}.
Let us find its radius $\kappa_s$.
A comoving mode will be excited if $\Lambda_k(\tau) \leq 1/\pi$.
Since $\Lambda_k(\tau) = {{\kappa^2} \over {a^2 \sqrt{q(\tau)}}}$,
we have 
\begin{equation}\label{cr1}
{{\kappa^2_s} \over {a^2 \sqrt{q(\tau)}}}
= a^{-1/2} {{\kappa^2_s} \over {\sqrt{q_0}}} \leq 1/\pi \ .
\end{equation}
Therefore for the light fermions, comoving radius of excited modes is
 increasing
with time as $\kappa_s \sim a^{1/4}$.
The radius of the sphere is scaled as
$\kappa_s \sim q_0^{1/4} a^{1/4}$.
% Figures of fermi sphere
%---------------------------------------------
%---------------------------------------------
% End of Figures
%---------------------------------------------
Due to the expansion, the amplitude 
 $\Phi$ is decreasing and the parameter $q(\tau)$ drops. 
At some moment the Fermi sphere will quit expansing.
Once  $q(\tau)$ is order of unity, the excitation is no longer strong,
 and redshifts of fermion modes will be fast enough to prevent from
 parametric excitation. Eventually fermions will be produced in the
perturbative regime.
If fermionic mass is non-zero, the perturbative decay goes unless
$2m_{\psi} > m_{\phi}$.

\subsection{\label{mass} Analytic Results for Production of Supermassive
 Fermions} 

One of the most significant differences between bosonic and fermionic parametric
 amplification is the possibility to create superheavy fermions from 
much liter inflatons, if inflatons are oscillating coherently.
As we  seen, for large $q$, massive  fermions are created not when
inflaton field itself crosses zero, but when
the combination $M_{eff}=h\phi+m_{\psi}$ crosses zero. This is because
at those instances even very heavy fermions are effectively massless,
because their bare mass is ``compensated'' by large value of $h\phi$,
if $\phi$ can have a large amplitude. This is exactly the  case
when $\phi$ is the inflaton field. For instance, in the chaotic
inflationary scenario amplitude of its oscillations immediately after inflation
 can be as large as $0.1 M_P$.

In this case we again have a situation when mode function of
heavy fermions has WKB form between zeros of $M_{eff}$, and can be described with
parabolic scattering around this instance.
Therfore our general formula  (\ref{nkfermi}) works for superheavy fermions as well.
The only modification is that the phase between successive scatterings
will be defined by the integral over time intervals
 between them.
In formula (\ref{nkfermi})  parameter $\Lambda_k(\tau)$ will be
\begin{equation}\label{lmass}
\Lambda_k(\tau)=-{{\pi \kappa^2} \over {a^2 \sqrt{{q_0 \over a^3} -
 \stackrel{\sim}{m}^2 } }}
\end{equation}
In case of massive fermions, the
criteria for excitation instead of (\ref{cr1}) will be 
\begin{equation}\label{cr2}
{{\kappa^2_s} \over {a^2 \sqrt{ {q_0 \over a^3} -
 \stackrel{\sim}{m}^2  }}} \leq 1/\pi \ .
\end{equation}
At the start of background oscillations for
large $q_0$ a term  $\stackrel{\sim}{m}^2$ is negligible.
Thus, as in the last section, we expect the Fermi sphere to fill with the
radius $\kappa_s=\sqrt{q_0}a^{1/4}$.
However, as  $ q$ drops, equation (\ref{cr2}) solved for 
$\kappa^2_s$ reaches a maximum, $\kappa_{m}$.
 and then decreases.
 This maximum occurs
when the scale factor is $a_m = \left( {q_0 \over {4 \stackrel{\sim}{m}^2}}
\right)^{1/3}$.  This gives $\kappa_{m}^2 =
 {\sqrt{3} \over \pi}\left({q_0 \over {4
 \sqrt{\stackrel{\sim}{m}}}}\right)^{2/3} $.
Notice that $k_m$ scales as $q_0^{1/3}$.
This is still not the final width of the band.
This is because there are a number of background oscillations 
which occurs between  the moment  $a_m$ and the moment when 
 denominator in (\ref{lmass})
approaches zero. We put a number of oscillation before  $a_m$
and between  $a_m$ and the moment of zero denominator as 
approximately equal and equal to $N_m$.
In this time after $a_m$ even the modes suppressed 
 exponentially  as $e^{-\Lambda_k}$ can achieve significant occupation numbers
in their random stochastic walk. We found the amplification factor
in the from of the exponent is  $\sqrt{N_m}$. 
In contrary to the ligh fermions case, there is no perturbative end
of the process. Excitation of superheavy
fermions is abruptly terminated when 
${q_0 \over a^3} -
 \stackrel{\sim}{m}^2$. 

\begin{figure}[tb]
\begin{minipage}[t]{8.0cm} %{7.2cm}
   \centering \leavevmode \epsfxsize=8.0cm    %7.2cm
   \epsfbox{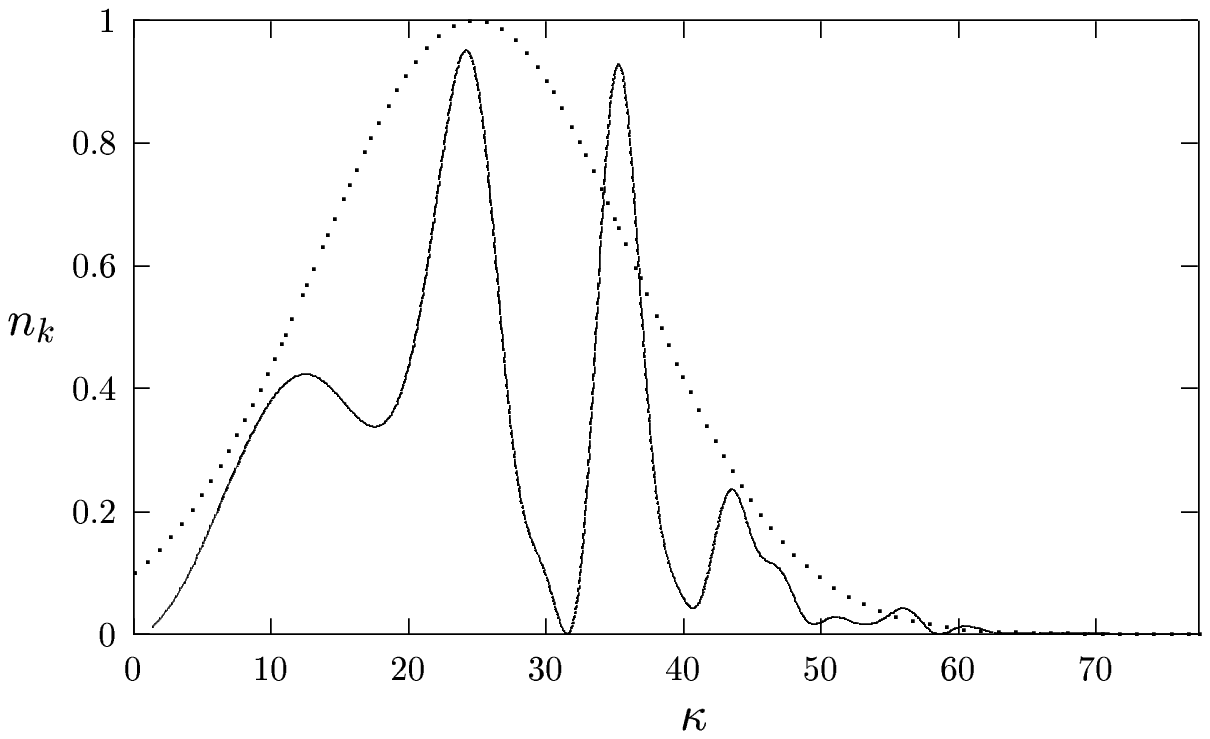}\\

   \caption[fig1u]{\label{fig:smf6} {\em 
Final spectrum of heavy fermions with masses $m_\psi = 50m_\phi$ when
$q_0 = 10^6$.  Also shown is the analytic estimate~(\ref{sup}).
\hspace*{\fill}}}
\end{minipage}
\hfill
\begin{minipage}[t]{8.0cm}
   \centering \leavevmode \epsfxsize=8.0cm
   \epsfbox{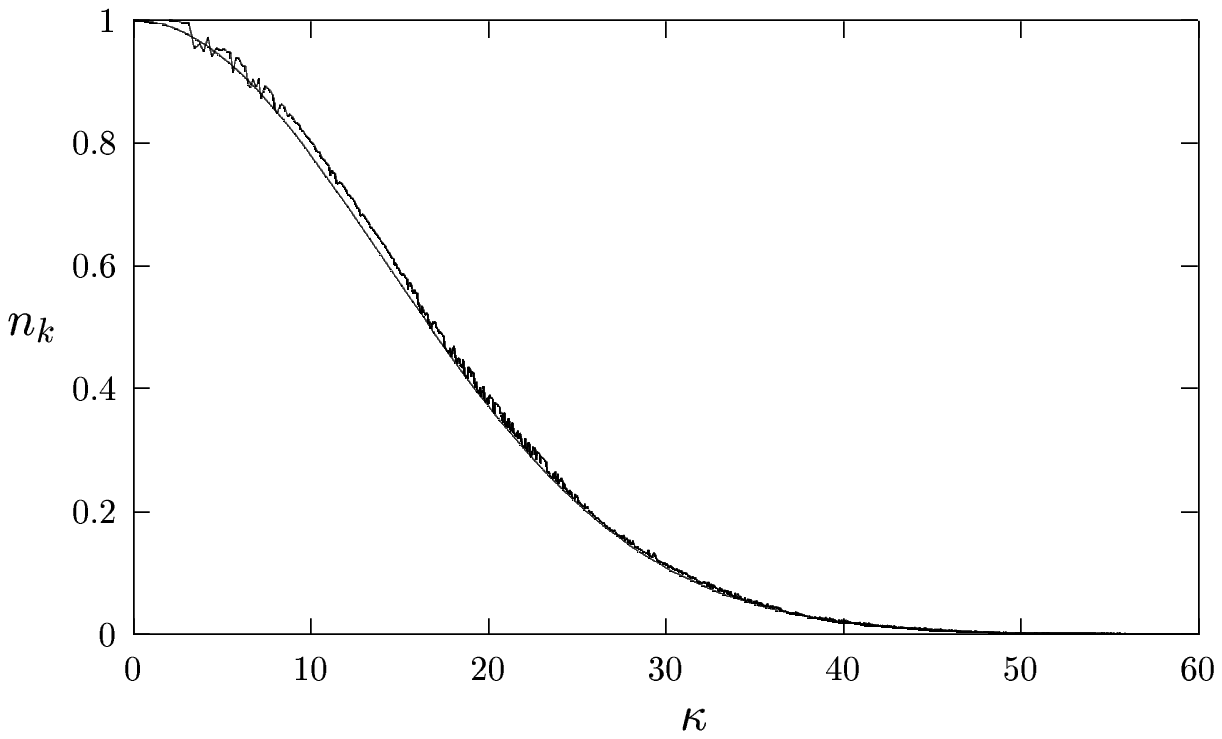}\\
   \caption[fig1u]{\label{fig:skick} {\em 
The comoving occupation number $n_k$ of fermions after $M_{eff}$  crosses one 
zero  for $q=10^6$ and $\stackrel{\sim}{m}=50$.
Also shown is the analytic formula~(35) which fits
perfectly.
  \hspace*{\fill}}}
\end{minipage}
\end{figure}

We found 
the spectrum of created superheavy particles after the process is terminated
 is given by
formula
\begin{equation}
n_k={1 \over 2 }\exp{\left(-2 { {  (\kappa-\gamma \kappa_m)^2  }\over 
  \kappa_m^2 } \right)}
\label{sup}
\end{equation}
where
$2\gamma^{2}= \ln \left( {q_0 \over
{4\pi^2 \stackrel{\sim}{m}^2}}
\right)$. This formula is valid for $N_m$ greater than a few.
In Fig.~\ref{fig:smf6} we plot numerically calculated final
spectrum of the supermassive fermions vs.  an analytic formula
\ref{sup}, which are in a  good agreement.
In the case of superheavy particles in an expanding universe
the maximum radius of the k-sphere is scaled with $q_0$ as
$\gamma k_m \sim q_0^{1/3}\ln q_0 $, which is  in 
close agreement with \cite{GPRT}.
However, one shall also check that at the moment when
excitation of superheavy fermions terminates, their back reaction
$h \langle \bar \psi \psi \rangle$  to the dynamics of inflaton oscillation
is still negligible.
% Figures of smf spectrum.
%---------------------------------------------
%---------------------------------------------
% End of Figures
%---------------------------------------------
In Fig.~\ref{fig:smf6} we plot spectrum of superheavy fermions
after the process is terminated.
Dashed curves are obtained with analytic formula (\ref{sup}).

\subsection{\label{broad} Fermionic Production from Single Kick}

As we seen formula (\ref{nkfermi}) can be extended to the case of expanding universe.
Consider the very first term in formula (\ref{nkfermi}). As
It corresponds to the creation
of fermions after effective mass $M_{eff}(t)$ crosses zero for the first time.
In this case we found
\begin{eqnarray}
n_k= e^{-\pi \Lambda_k^2}
\label{kick}
\end{eqnarray}
Occupation number of fermions generated from the single kick is plotted
in Fig.~\ref{fig:skick}. Two curves corresponding to the formula 
(\ref{kick}) and to the numerical solutions are shown and practically
 indistinguishable.

\section{\label{sum} Discussion } 

In this paper we developed  a non-perturbative theory for fermion 
production by an oscillating inflaton  field. As we have  seen, the 
production of fermions can be characterized as parametric excitation.
Even in the simple model of a Yukawa coupling between fermions and
a background scalar field in an expanding universe
oscillating around the minimum of its quadratic potential, 
the theory of fermionic parametric excitation is rich and leads to
important results.  Fermions are created very quickly, within about ten(s)
oscillations, in out-of-equilibrium states. 
For large values of the resonance parameter $q={{h\phi_0} \over m_\phi}$,
occupation number of light fermions ($m_\psi < m_\phi \ll h\phi_0$)
is changing in a step-like manner 
 at instances $t_j$ when the  inflaton  amplitude $\phi(t_j)=0$ passes
through zero, $j=1,2,3, ..$
We have developed the method of parabolic scatterings for fermions, 
based closely  on a similar approach for the bosonic resonance \cite{KLS97}.
It is possible to derive a unified recursive formula, which relates
the occupation number of  fermions or bosons
$n^{j+1}_k$ at the moment $t_{j+1}$ to the earlier value $n^{j}_k$ :
\begin{eqnarray}
n^{j+1}_k &=& e^{-\pi \Lambda_k^2} + \left( 1 \pm  2 e^{-\pi \Lambda_k^2}
\right) n^{j}_k\nonumber \\ &-&
2 (-1)^j e^{-{\pi \over 2} \Lambda_k^2}\sqrt{1 \pm e^{-\pi \Lambda_k^2}}
\sqrt{ n^{j}_k (1 \pm n^{j}_k)} \sin \theta_{tot}^{j} \ ,
\label{nkanal}
\end{eqnarray}
For bosons one shall use an upper sign and neglect the $(-1)^j$
terms~\cite{KLS97}, while for fermions 
one shall use the lower sign. For light fermions and bosons
 $\Lambda_k^2={k^2 \over {\sqrt{q} m_\phi^2}}$.
For large $q$ the angle  $\theta_{tot}^{j}$ can be treated as a random phase. 
As a result, formula (\ref{nkanal}) predicts the  
stochastic character of parametric excitation of both bosons and fermions.
In the fermionic case it leads to the conclusion that 
in the momentum space fermions chaotically fill up a broad sphere of the radius
$\sim q^{1/4}m$.  This formula also clearly shows the features of:
spontaneous emission: for both bosons and fermions, the first
oscillation leads to the spectrum $n_k = e^{-{\pi \Lambda_k}}$;
stimulated emission  for bosons with $n_k \gg 1$, we have 
$(n_k^{j+1}- n_k^j)  \propto n_k^j$;
Pauli Blocking for fermions, if $n_k^j =1$, the next value will always
be $n_k^{j+1} = 1 - e^{-{\pi \Lambda_k}}$ even in the stochastic case.  This
prevents the occupation number of fermions from exceeding $1$.

Formula (\ref{nkanal}) can be extended to the case of massive bosons and
fermions.
However, here important differences emerge. For bosons we will have
 $\Lambda_k={{k^2+m_{b}^2} \over {\sqrt{q}m_\phi}}$, where
$m_{b}$ is the $\chi$-boson mass. It leads to the conclusion that the
creation of superheavy, $m_b > m_\phi$,
bosons is exponentially suppressed.
However, for fermions we have $\Lambda_k={k^2 \over 
{m_\phi^2 sqrt{h^2\phi_0^2-m_{\psi}^2}}} $.
Therefore even superheavy fermions with the mass as large as $h\phi_0$
can be created in abundance from the coherent inflaton oscillations
\cite{Cosmo,GPRT}.
This occurs because the effective mass of fermions is given by the
algebraic combination $h\phi(t)+m_{\psi}$ and the creation of fermions
occurs when the effective mass goes to zero. The method of parabolic
scatterings, applied for instances when this happens,
leads to both formulae $(\ref{nkanal})$ with the corresponding $\Lambda_k$.

There are situations where single instance (single kick) of particle creation 
may lead to interesting effects. An example is  the scenario of instant
preheating, which is especially important for inflationary models
without minima of the inflaton potential \cite{inst}. Another example is
the interaction of the inflaton with
superheavy fermions during the inflationary stage  when the combination
$h\phi(t)+m_{\psi}$ can go through zero only once.
The generic formula (\ref{nkanal}) embraces 
 the case where  bosons  and fermions are created by a single kick. We 
shall put
$j=0$, $n_k^0=0$,
and then $n^{1}_k = e^{-\pi \Lambda_k^2}$. However, instead of the 
parameter $q$ we shall use another combination of coupling constants and
$\phi_0$ and $m_\phi$. In this case the effect is defined by the
velocity $\dot \phi_*$ at the moment $t_*$, which is different for bosons
and fermions.
For bosons single instance  creation of particles gives the spectrum
  \cite{KLS97,inst}
\begin{eqnarray}
n_k= e^{-{{\pi (k^2+m_{\chi^2})} \over {g \dot \phi_*}}}
\label{bo}
\end{eqnarray}
where $t_*$ corresponds to $\phi(t_*)=0$.
For fermions single instance  creation gives \cite{infl}
\begin{eqnarray}
n_k= e^{-{{\pi k^2} \over {h \dot \phi_*}}}
\label{fe}
\end{eqnarray}
where $\phi(t_*)+m_{\psi}/h =0$.

We believe  that the theory of fermionic preheating 
will be important ingredient of the realistic scenarios of the reheating of the
universe after inflation.\\

{\bf Acknowledments.} 
Collaboration and discussions with 
Andrei Linde and Alexei Starobinsky significantly influenced 
this work. We also thank J\"{u}gen Baacke for useful discussion.
This work was supported by NSERC and CIAR.

\end{document}